\documentclass[12pt]{article}
\usepackage{epsfig}
\usepackage{url}
\makeatletter
\def\alt{\mathrel{\mathpalette\gl@align<}}
\def\agt{\mathrel{\mathpalette\gl@align>}}
\def\gl@align#1#2{\lower.6ex\vbox{\baselineskip\z@skip\lineskip\z@
\ialign{$\m@th#1\hfil##\hfil$\crcr#2\crcr\sim\crcr}}}
\makeatother

\begin{document}
\begin{flushright}
%{\tt hep-ph/yymmnnn}\\
MIFP-09-04 \\
January, 2009
\end{flushright}
\vspace*{2cm}
\begin{center}
\baselineskip 20pt
{\Large\bf
Penguin Contribution to the Phase of $B_s$-$\bar B_s$ Mixing \\
and
$B_s \to \mu \mu$
in Grand Unified Theories
} \vspace{1cm}

{\large
Bhaskar Dutta and Yukihiro Mimura}
\vspace{.5cm}

{\it
Department of Physics, Texas A\&M University,
College Station, TX 77843-4242, USA
}
\vspace{.5cm}

\vspace{1.5cm}
{\bf Abstract}
\end{center}
We investigate the possibility of a large $B_s$-$\bar{B_s}$ mixing phase in the context of grand unified theory (GUT) models, e.g., SO(10) and SU(5). In these models, we find that a large phase of $B_s$ mixing is correlated with Br($b\to s\gamma$), Br($\tau\to\mu\gamma$) and Br($B_s \to\mu\mu$) for large $\tan\beta$. In the case of the SO(10) model, the large phase of $B_s$ mixing is correlated with Br($b\to s\gamma$) and Br($B_s \to\mu\mu$) and we find that a large $B_s$ mixing corresponds to an enhanced Br($B_s \to\mu\mu$) about to be probed by the Tevatron. In the case of the SU(5) model, the large phase is correlated with Br($\tau\to\mu\gamma$) and Br($B_s \to\mu\mu$). In this case, the Br($\tau\to\mu\gamma$) constraint requires a smaller pseudo-scalar Higgs mass which in turn generates a large Br($B_s \to\mu\mu$) almost at the edge of present experimental constraint. If the present observation of large phase of $B_s$ mixing persists in the upcoming data, using all these branching ratios, we will be able to distinguish these models.

\thispagestyle{empty}

\bigskip
\newpage

\addtocounter{page}{-1}

\section{Introduction}
\baselineskip 18pt

Recently, CDF and D$\O$ collaborations
have announced the analysis
of the flavor-tagged $B_s \to J/\psi \phi$ decay. The decay width difference and the mixing induced
CP violating phase, $\phi_s$, were extracted from their analysis \cite{Aaltonen:2007he}.
In the Standard Model (SM),
the CP violating phase is predicted to be
small,
$\phi_s = 2\beta_s \equiv 2\, {\rm arg}\, (-V_{ts}V_{tb}^*/V_{cs}V_{cb}^*) \simeq 0.04$.
However, the measurements of the phase are large:
\begin{eqnarray}
\phi_s ({\rm CDF}) &\in& [0.28,1.29]\ \ (68\%\, {\rm C.L.}), \\
\phi_s ({\rm D\O}) &=& 0.57^{+0.30}_{-0.24}({\rm stat})
                       {}^{+0.02}_{-0.07}({\rm syst}).
\end{eqnarray}
The UTfit group made a combined data analysis
including the semileptonic asymmetry in the $B_s$ decay,
and find that the CP violating phase deviates more
than $2.5 \sigma$ from the SM prediction \cite{Bona:2008jn}.
%
%This implies the existence of new physics (NP) and
%that the NP model requires a flavor violation in $b$-$s$ transition.
%
If this large phase still persists in the upcoming results from Fermilab,
it implies the existence of new physics (NP) beyond SM
and that the NP model requires a flavor violation in $b$-$s$ transition
as well as a phase in the transition.

The nature of flavor changing neutral currents (FCNCs)
and the CP violating phase is very important to test the existence of new physics %(NP)
beyond the standard model. %(SM).
Supersymmetry (SUSY) is the most attractive candidate to build NP models.
The gauge hierarchy problem can be solved and a natural aspect of
the theory  can be developed from the weak scale to the ultra high
energy scale.
In fact, the gauge coupling constants in the standard model %(SM)
can unify at a high scale using the  renormalization group equations
(RGEs) involving the particle contents of the minimal SUSY standard
model (MSSM), which indicates the existence of grand unified
theories (GUTs).
 The well motivated SUSY GUTs %grand unified theories (GUT)
have always been subjects of intense experimental and theoretical investigations.
Identifying a GUT model will be a major focus of the upcoming experiments.

%Also, the SUSY standard model can provide a candidate to the dark matter
%even in the minimal version of the model.

%Supersymmetry (SUSY) is the most attractive candidate to construct the NP models.
%
In SUSY models, the SUSY breaking mass terms for squarks and sleptons
must be introduced, and they have sources of FCNCs and CP violation beyond the Kobayashi-Maskawa
theory.
In general, they generate too large FCNCs,
and thus
the flavor universality is often assumed
in squark and slepton mass matrices
to avoid the large FCNCs  %flavor changing neutral currents (FCNC)
in the meson mixings and the lepton flavor violations (LFV) \cite{Gabbiani:1988rb}.
The flavor universality is expected to be realized by the Planck scale physics.
However,
even if the flavor universality is realized at a scale such as the GUT scale
or the Planck scale,
the non-universality in the SUSY breaking sfermion masses is generated
from the evolution of RGEs, %renormalization group equations (RGE).
and they can generate a small flavor violating transitions,
which can be observed in the ongoing experiments.

In the MSSM %the minimal extension of SUSY standard model (MSSM)
with right-handed neutrinos,
the induced FCNCs from RGE effects are not large in the quark sector,
while sizable effects can be generated in the lepton sector due to
the large neutrino mixing angles \cite{Borzumati:1986qx}.
In GUTs, %the grand unified theories (GUT),
the loop effects due to the large neutrino mixings
can also induce sizable effects in the quark sector
since GUT scale particles can propagate in the loops~\cite{Barbieri:1994pv}.
As a result, the patterns of the induced FCNCs
highly depend on the unification scenario,
and the contents of the heavy particles.
Therefore, it is important to investigate
the FCNC effects to obtain a footprint of the GUT models.
If the quark-lepton unification is manifested
in GUT models,
the flavor violation in $b$-$s$ transition
can be responsible for the large atmospheric neutrino mixing \cite{Moroi:2000tk},
and
thus, the amount of the flavor violation in $b$-$s$ transition
(the second and the third generation mixing),
which is related to the $B_s$-$\bar B_s$ mixing and its phase,
 has to be related to the $\tau \to \mu\gamma$ decay \cite{Dutta:2006gq,Parry:2005fp,Dutta:2008xg,Hisano:2008df}
for a given particle spectrum.
The branching ratio of the $\tau \to \mu\gamma$
is being measured at the $B$-factory,
and thus, the future results of LFV and the ongoing measurement of the
phase of $B_s$-$\bar B_s$ mixing
will provide an important information to probe
the GUT scale physics.

In Ref.\cite{Dutta:2008xg},
we have studied the correlation between Br($\tau\to\mu\gamma$)
and $\phi_s$, the phase in $B_s$-$\bar B_s$ mixing,
comparing SU(5) and SO(10) GUT models,
and investigated the constraints in these models from the observations
in order to decipher GUT models.
The flavor violation originating from the
loop correction via the heavy particles
can be characterized by the
CKM (Cabibbo-Kobayashi-Maskawa) quark mixing matrix and the
MNSP (Maki-Nakagawa-Sakata-Pontecorvo) neutrino mixing matrix,
as well as the size of the Yukawa couplings.
Since the CKM mixings are small, it is expected that the neutrino mixings
 dominate the source of FCNCs at low energy.
It is important to know whether
the large neutrino mixings originate from the Dirac
neutrino Yukawa coupling or the Majorana-type Yukawa coupling.
When the large neutrino mixings originate from the Dirac
neutrino Yukawa couplings in a GUT model,
the (squared) right-handed down-type squark mass matrix, $M_{\tilde D^c}^2$,
as well as the left-handed lepton doublet mass matrix, $M_{\tilde L}^2$,
 can
have  flavor non-universality.
When the large mixings originate from the Majorana Yukawa couplings,
the left-handed squark mass matrix, $M_{\tilde Q}^2$,
can also have  flavor non-universality in addition to the other sfermions.

In the minimal-type of SU(5) GUT,
the large neutrino mixing originates from the Dirac neutrino coupling
if there is no fine-tuning in the seesaw neutrino matrix.
On the other hand, in the minimal-type of SO(10) GUT,
the large neutrino mixing can originate from the Majorana-type coupling.
%
%Actually, since the right-handed neutrino is unified to the other
%fermion species in the SO(10) GUT, the Majorana coupling
%can be unified to the fermion
%
In general, since SU(5) is a subgroup of SO(10),
one can construct a model where the neutrino mixing originate from
the Majorana-type coupling in non-minimal-type of SU(5) GUT.
Also, if we allow the fine-tuning in the Yukawa coupling matrices,
the Dirac neutrino Yukawa coupling can be the source of the
large mixing even in the SO(10) model.
Actually, there is a little ambiguity to determine the minimal SU(5) or SO(10) GUT model,
since minimal versions of the GUT models have problems with phenomenology.
(That is why we call them minimal-type.)
Here, we call the typical boundary condition as minimal-type of SU(5) GUT condition
when the off-diagonal elements of $M_{\tilde D^c}^2$ and $M_{\tilde L}^2$
are correlated due to the Dirac neutrino coupling in GUT models.
The other boundary condition where the $M_{\tilde Q}^2$ is also correlated to $M_{\tilde D^c, \tilde U^c}^2$ and $M_{\tilde L}^2$ due
to the Majorana coupling in SO(10) model is called as minimal-type of SO(10) GUT
boundary condition.
The large phase of $B_s$-$\bar B_s$ mixing, as well as the other flavor violating processes,
can tell us which type of boundary condition is preferable.

%In the minimal-type of SU(5) model,
%the flavor violation originates from the Dirac neutrino Yukawa coupling.

We analyzed the case of lower $\tan\beta$
(which is a ratio of the vacuum expectation values of up- and down-type Higgs fields)
in the Ref.\cite{Dutta:2008xg}.
In such a case, the box diagram contribution will dominate the SUSY contribution
of $B_s$-$\bar B_s$ mixing amplitude,
and we found that the SO(10) boundary condition is more important to obtain the
large phase of $B_s$-$\bar B_s$ mixing.
When $\tan\beta$ is large, the so called double penguin contribution \cite{Hamzaoui:1998nu,Buras:2001mb}
can dominate the SUSY contribution rather than the box contribution
unless the pseudo Higgs field is heavy.
In such cases,
the $B_s \to \mu\mu$ decay \cite{Choudhury:1998ze,Buras:2001mb}
will be enhanced close to its experimental bound \cite{Foster:2004vp}.
In other words, if the large phase of $B_s$-$\bar B_s$ mixing
originates from the double penguin contribution,
the $B_s \to \mu\mu$ decay will be observed very soon,
and it is worth to examine the constraints if
a large phase is really generated from the double penguin contribution.
In this paper, we will investigate the double penguin contribution
of the $B_s$-$\bar B_s$ mixing,
as well as the other flavor violating processes
including $B_s \to \mu\mu$, $b\to s\gamma$, and $\tau\to\mu\gamma$ in the context of SO(10) and SU(5) models.

The paper is organized as follows:
In section 2, we will describe the FCNC sources in SUSY GUT models.
The two typical boundary conditions in both SU(5) and SO(10) model
are considered.
In section 3, we will describe the SUSY contributions of $B_s$-$\bar B_s$ mixing amplitudes,
including the box diagram and the double penguin contribution.
The constraint from $B_s \to \mu\mu$, $b\to s\gamma$, $\tau\to\mu\gamma$ in the models
are also noted.
In section 4, we will show our numerical work on the both kinds of the GUT models.
Section 5 devotes the conclusion and remarks.

\section{FCNC sources in SUSY GUTs}

In SUSY theories, the SUSY breaking terms can be the sources of
 flavor violations.
In general, it is easy to include  sources of flavor
violation  by hand
 since the SUSY
breaking masses with flavor indices are parameters in the MSSM.
However, if these parameters are completely general, too much FCNCs
are induced \cite{Gabbiani:1988rb}.
Therefore, as a minimal assumption of the SUSY breaking,
  the universality of scalar masses is often considered,
which means that all the SUSY breaking (squared) scalar masses are universal
to be $m_0^2$, and the scalar trilinear couplings are proportional
to Yukawa couplings (the coefficient is universal to be $A_0$) at a
unification scale.
Even if the universality is assumed,
the non-universality in scalar masses is generated from the evolution of the theory
from the GUT scale down to the weak scale via RGEs.
In the MSSM with right-handed neutrino ($N^c$), the induced FCNCs from RGE effects
are not large in the quark sector, while sizable effects can be generated
in the lepton sector due to the large neutrino mixings \cite{Borzumati:1986qx}.
The sources of FCNCs in the model are the Dirac neutrino couplings.

In GUT models, the left-handed lepton doublet ($L$) and the right-handed down-type squarks ($D^c$)
are unified in $\bar{\bf 5}$, and
the Dirac neutrino couplings can be written as $Y_\nu \bar{\bf 5} N^c H_{\bf 5}$.
As a result, non-universality in the SUSY breaking mass matrix for $D^c$
is also generated from the colored-Higgs and right-handed neutrino loop diagram,
and the flavor violation in the quark sector can be generated from the Dirac neutrino
couplings \cite{Barbieri:1994pv,Moroi:2000tk}.

The light neutrino mass matrix is written as
\begin{equation}
{\cal M}_\nu^{\rm light} = f \langle \Delta_L \rangle - Y_\nu M_R^{-1} Y_\nu^{\rm T} \langle H_u^0 \rangle^2,
\end{equation}
where $\Delta_L$ is an SU(2)$_L$ triplet, and $f$ is a Majorana coupling $\frac12 LL\Delta_L$.
The second term is called type I seesaw term \cite{Minkowski:1977sc}.
If the type I seesaw term dominates the light neutrino mass,
the Dirac neutrino coupling will have large mixings to explain the large neutrino mixings
in the basis where the charge-lepton Yukawa coupling $Y_e$ is diagonal.
On the other hand, when the first term (triplet term) dominates it (type II seesaw \cite{Schechter:1980gr}),
the Majorana coupling must have the large mixings.
Distinguishing these two cases is very important in order to understand the source of FCNCs in the GUT models.

Let us first describe the non-universality from the Dirac neutrino couplings.
We will work in a basis where the charged-lepton Yukawa matrix, $Y_e$,
and the right-handed neutrino Majorana mass matrix, $M_R$, are
diagonal,
\begin{equation}
M_R = {\rm diag}\, (M_1,M_2,M_3).
\end{equation}
The neutrino Dirac Yukawa coupling matrix is written as
\begin{equation}
Y_\nu = U_L Y_\nu^{\rm diag} U_R^{\rm T},
\label{neutrino-Dirac}
\end{equation}
where $U_{L,R}$ are diagonalizing unitary matrices.
We note that $U_L$ corresponds to the (conjugate of)
MNSP neutrino mixing matrix, $U_{\rm MNSP}$,
in type I seesaw,
%$m_\nu^{\rm light} = Y_\nu M_R^{-1} Y_\nu^{\rm T}\langle H_u^0 \rangle^2$ \cite{Minkowski:1977sc},
up to a diagonal phase matrix if $U_R$ is exactly same as ${\bf 1}$ (identity matrix),
which we will assume for simplicity.
Through RGEs, the off-diagonal elements of
the SUSY breaking mass matrix for the left-handed lepton doublet
gets the following correction
\begin{equation}
\delta M_{\tilde L}^2{}_{ij} \simeq -\frac{1}{8\pi^2} (3 m_0^2 + A_0^2)\,
\sum_k (Y_\nu)_{ik} (Y_\nu^*)_{jk} \ln \frac{M_{*}}{M_k},
\end{equation}
where $M_*$ is a cutoff scale and
the SUSY breaking parameters are universal.
Neglecting the threshold of the GUT and the Majorana mass scales,
we can write down the boundary conditions as
\begin{equation}
M_{\bar{\bf 5}}^2 =
M_{\tilde D^c}^2 = M_{\tilde L}^2 = m_0^2
\left(
{\bf 1} - \kappa\, U_L
\left(
\begin{array}{ccc}
k_1 & & \\
& k_2 & \\
&& 1
\end{array}
\right) U_L^{\dagger}
\right),
\label{boundary-5}
\end{equation}
where $\kappa
\simeq (Y_\nu^{\rm diag})_{33}^2
(3+A_0^2/m_0^2)/8\pi^2 \ln M_*/M_{\rm GUT}$, and
$k_2 \simeq \sqrt{\Delta m_{\rm sol}^2/\Delta m_{\rm atm}^2}
M_2/M_3$.
%when $V_R^e \simeq {\bf 1}$.
%
We parameterize the unitary matrix $U_L$ as
\begin{equation}
 U_L =
\left(\begin{array}{ccc}
e^{i (\alpha_1 - \delta)} & & \\
& e^{i \alpha_2} & \\
& & 1
\end{array}
\right)
\left(
\begin{array}{ccc}
%c^e_{12} c^e_{13} & -s^e_{12} c^e_{23} - c^e_{12}s^e_{23} s^e_{13}e^{-i \delta} &
%s^e_{12} s^e_{23}- c^e_{12} c^e_{23} s^e_{13} e^{-i \delta}\\
%s^e_{12} c^e_{13} & - s^e_{12} s^e_{13} s^e_{23} + c^e_{12} c^e_{23}e^{-i \delta}
%& -c^e_{12} s^e_{23} -s^e_{12} s^e_{13} c^e_{23} e^{-i \delta}\\
%s^e_{13}e^{i \delta} & c^e_{13} s^e_{23} & c^e_{13} c^e_{23}
c^e_{12} c^e_{13} & s^e_{12} c^e_{13} & s^e_{13}e^{i \delta} \\
-s^e_{12} c^e_{23} - c^e_{12}s^e_{23} s^e_{13}e^{-i \delta} &
c^e_{12} c^e_{23} - s^e_{12} s^e_{13} s^e_{23} e^{-i \delta} &
c^e_{13} s^e_{23} \\
s^e_{12} s^e_{23}- c^e_{12} c^e_{23} s^e_{13} e^{-i \delta} &
-c^e_{12} s^e_{23} -s^e_{12} s^e_{13} c^e_{23} e^{-i \delta} &
c^e_{13} c^e_{23}
\end{array}
\right),
\label{MNSP}
\end{equation}
where $s_{ij}^e$ and $c_{ij}^e$ are sin and cos of mixing angles
$\theta_{ij}$. In the limit $k_{1,2} \to 0$, $\alpha_1$ and
$\alpha_2$ are the phases of the 13 and the 23 element of $M_{\bar
{\bf 5}}^2$.
Since we are assuming that $U_R = \bf 1$, $\theta_{12}$ and
$\theta_{23}$ correspond to solar and atmospheric neutrino mixings,
respectively, which are large.
Even if we do not assume $U_R = \bf 1$,
the angle $\theta_{23}$ is expected to be large
unless there exists a fine-tuned relation among $Y_\nu^{\rm diag}$ and $M_N$.
Assuming that the Dirac neutrino Yukawa coupling is hierarchical ($k_1, k_2 \ll 1$),
we obtain the 23 element of $M_{\bar {\bf 5}}^2$ as
$-1/2 \,m_0^2\, \kappa \sin 2\theta_{23}\, e^{i\alpha_2}$.
Therefore, the magnitude of the FCNC between 2nd and 3rd generations
is controlled by $\kappa \sin2\theta_{23}$.
The phase $\alpha_2$ will be the origin of a phase of SUSY contribution of
$B_s$-$\bar B_s$ mixing amplitude.
%
%where $\theta_{23}$ is a 2-3 mixing in $U_L$,
%which is large and responsible for the large atmospheric neutrino mixing
%unless there exists a fine-tuned relation among $Y_\nu^{\rm diag}$ and $M_N$.
%The phase $e^{i \alpha}$ generates a phase of SUSY contribution for
%$B_s$-$\bar B_s$ mixing amplitude, $M_{12}^{\rm SUSY}$,
%and the absolute value of $M_{12}^{\rm SUSY}$ is controlled by $\kappa \sin 2\theta_{23}$.
%
%
%On the other hand, $s_{13}^e$ is
%bounded by CHOOZ experiments %, $s_{13}^e \alt 0.2$
%\cite{Apollonio:1999ae}. %
%
The SUSY breaking mass for $\bf 10$
multiplet $(Q,U^c,E^c)$ is also corrected by the (colored-)Higgsino loop, but
it arises from CKM mixings and the effect is small.
So, the boundary condition at the GUT scale for $\bf 10$ multiplet is
\begin{equation}
M_{\bf 10}^2 = M_{\tilde Q}^2 = M_{\tilde U^c}^2 = M_{\tilde E^c}^2
\simeq m_0^2 \,{\bf 1}.
\label{boundary-10}
\end{equation}
%neglecting the small contribution  arising from the CKM mixings.
%
The boundary conditions, Eqs.(\ref{boundary-5},\ref{boundary-10}),
are the typical boundary conditions in the case of minimal kind of SU(5) GUT
with type I seesaw \cite{Dutta:2008xg,Dutta:2006zt}.

The Yukawa coupling matrices for up- and down-type quarks and
charged-leptons are given as
%
%The minimal SU(5) boundary condition is the following:
%
\begin{eqnarray}
Y_u &=& V_{L} V_{\rm CKM}^{\rm T} Y_u^{\rm diag} P_u V_{uR}^{\rm T},
\label{Yukawa-u} \\
Y_d &=& V_{L} Y_d^{\rm diag} P_d V_{dR}^{\rm T},
\label{Yukawa-d} \\
Y_e &=& Y_e^{\rm diag} P_e,
\label{Yukawa-e}
\end{eqnarray}
where $Y_{u,d,e}^{\rm diag}$ are real
(positive) diagonal matrices and $P_{u,d,e}$ are diagonal phase
matrices.
In the minimal SU(5) GUT, in which only $H_{\bf 5}$ and
$H_{\bar {\bf 5}}$ couple to matter fields,
we have $V_{uR} = V_{\rm CKM}^{\rm T}$,
$V_{L} = V_{R} = {\bf 1}$,
and $Y_d^{\rm diag} = Y_e^{\rm diag}$.
Because it will give us a wrong prediction to the quark and charged-lepton masses,
we need at least a slight modification from the minimal assumption.
Even if there is a slight modification of the Yukawa coupling,
we assume that the unitary matrix $V_{dR}$ does not have large mixings.
If $V_{dR}$ has a large mixing, the FCNC sources in $M_{\tilde D^c}^2$
may be cancelled in the basis where the down-type quark mass matrix is
diagonal.

Next, let us consider the case of type II seesaw in the framework of SO(10) GUT
models \cite{Babu:1992ia,Dutta:2004wv}.
All matter fields are unified in the spinor representation $\bf 16$
in the
 SO(10) models.
Since the right-handed neutrino is also unified to other matter fields, the
 neutrino Dirac Yukawa coupling does not
have large mixings ({\it i.e.} $U_{L} \simeq {\bf 1}$)
if there is no large cancellation in the Yukawa couplings.
%in a simple fit of the Yukawa couplings.
%
In this case, as we have mentioned,
the proper neutrino masses with large mixings
can be generated from the %type II seesaw mechanism \cite{Schechter:1980gr},%in which the
Majorana couplings $\frac12 f LL \Delta_L$. %(where $\Delta_L$ is a SU(2)$_L$ triplet)
% induces the light neutrino masses,
%$m_\nu^{\rm light} = f \langle \Delta_L^0 \rangle$.
%
Due to the
unification under SO(10), the left-handed Majorana coupling, $f$, is
tied to all other matter fields, and therefore, the off-diagonal terms in
the sparticle masses are induced by loop effect which are
proportional to $f f^\dagger$. Neglecting the GUT scale threshold,
we can write the boundary condition in SO(10) as
\begin{equation}
M_{{\bf 16}}^2 =
m_0^2
\left(
{\bf 1} - \kappa \, U
\left(
\begin{array}{ccc}
k_1 & & \\
& k_2 & \\
&& 1
\end{array}
\right) U^\dagger
\right),
\label{boundary-16}
\end{equation}
where $\kappa \simeq 15/4\,(f_{33}^{\rm diag})^2
(3+A_0^2/m_0^2)/8\pi^2 \ln M_*/M_{\rm GUT}$,
and $k_2 \simeq \Delta m_{\rm sol}^2/\Delta m_{\rm atm}^2$
in this case.
Note that the parameters $\kappa$, $k_{1,2}$ are of course
different from those given in Eq.(\ref{boundary-5})
using the set-up for type I seesaw,
but we use the same notation to simplify the description.
The unitary matrix $U$ is the (conjugate of) MNSP neutrino mixing
matrix up to a diagonal phase matrix, which is parameterized in the
same way as Eq.(\ref{MNSP}). The Yukawa couplings are also given as
Eq.(\ref{Yukawa-u},\ref{Yukawa-d},\ref{Yukawa-e}). If we do not
employ $\bf 120$ Higgs fields, the Yukawa matrices are symmetric, and
thus, $V_{uR} = V_L V_{\rm CKM}^{\rm T}$, $V_{dR}= V_{L}$. The
unitary matrix $V_{L}$ is expected to be close to $\bf 1$ if there
is no huge fine-tuning in the fermion mass fits.

Note that the sources of phases are not only in the unitary matrix $U$
but also in the phase matrix $P_d$ in the Yukawa coupling.
Actually, in the basis where down-type quark mass matrix is a real (positive) diagonal matrix,
the phases of the 23 elements in $M_{\tilde Q}^2$ and $M_{\tilde D^c}^2$ are
independent,
and these are two independent phase parameters which act as FCNC and CP violating sources for $b$ to $s$ transition.

If the SO(10) symmetry is manifested above the GUT symmetry breaking threshold,
the off-diagonal elements of SUSY breaking sfermion mass matrices are
unified at the GUT scale.
However, depending on a Higgs spectrum, the symmetry breaking
of the SO(10) symmetry may not happen at a single scale.
Actually, the Higgs spectrum from $\bf 126$ Higgs can be split
depending on a vacuum of the SO(10) symmetry breaking.
At that time, the magnitude of the off-diagonal elements depends on the
sfermion species:
\begin{equation}
M_{\tilde {F}}^2 = m_{0}^2 [{\bf 1} - \kappa_F U {\rm diag} (k_1,k_2,1) U^{\dagger}],
\end{equation}
where $F = Q, U^c, D^c, L, E^c$.
The quantity $\kappa_F$ denotes the amount of the off-diagonal elements
and it depends on the sfermion species.
For example, only uncolored GUT particles are light compared to the others
as a result of the SO(10) breaking,
$\kappa_L$ and $\kappa_{E^c}$ are larger than the others,
and the lepton flavor violation will be enhanced rather than the quark flavor violation.

It is interesting that the flavor violation pattern in the lepton sector
and the quark sector can depend on the SO(10) symmetry breaking vacua.
Actually, in order to forbid a rapid proton decay,
the quark flavor violation should be larger than the lepton flavor violation
among the symmetry breaking vacua \cite{Dutta:2007ai}.
Namely, it is expected that
$\kappa_{Q}$, $\kappa_{U^c}$, and $\kappa_{D^c}$ are much larger than
$\kappa_L$ and $\kappa_{E^c}$.
For example, if only the Higgs fields $({\bf 8},{\bf 2}, \pm1/2)$
are light compared to the breaking scale (which is the most suitable case),
one obtains $\kappa_Q = \kappa_{U^c} = \kappa_{D^c}$,
and only quark flavor violation is generated, while lepton flavor violation is not.
On the other hand, when the flavor violation is generated
from the minimal-type of SU(5) vacua with type I seesaw,
the quantities $\kappa$'s have relations as $\kappa_{L} \sim \kappa_{D^c}$, and
$\kappa_Q, \kappa_{U^c},\kappa_{E^c} \sim 0$, effectively.
Actually, when we take into account the threshold effect,
it is expected that $\kappa_{L}$ is always larger than $\kappa_{D^c}$
since the right-handed Majorana mass scale is less than the scale of colored Higgs
mass.
Therefore, the existence of $b$-$s$ transition indicated by the experimental results
in Fermilab predicts the sizable lepton flavor violation in the minimal-type of SU(5) model.
Thus, if the results of large $B_s$-$\bar B_s$ phase is really an evidence of NP,
the GUT models are restricted severely \cite{Parry:2005fp,Dutta:2008xg,Hisano:2008df}.

Therefore, investigating the quark and lepton flavor violation is very important to decipher
the GUT symmetry breaking, when the $B_s$-$\bar B_s$ phase is large~\cite{Dutta:2008xg}.

\section{$B_s$-$\bar B_s$ mixing and the other flavor violating processes}

Let us briefly see the phase of $B_s$-$\bar B_s$ mixing.
We use the model-independent parameterization of the NP contribution:
\begin{equation}
C_{B_s} e^{2i\phi_{B_s}} = M_{12}^{\rm full}/M_{12}^{\rm SM},
\end{equation}
%
%\begin{equation}
%$C_{B_s} e^{2i\phi_{B_s}} = M_{12}^{\rm full}/M_{12}^{\rm SM}$,
%\end{equation}
where `full' means the SM plus NP contribution,
$M_{12}^{\rm full} = M_{12}^{\rm SM}+ M_{12}^{\rm NP}$.
The NP contribution can be parameterized by two real parameters $C_{B_s}$
and $\phi_{B_s}$.
The time dependent CP asymmetry ($S = \sin \phi_s$) in $B_s \to J/\psi \phi$
is dictated by the argument of $M_{12}^{\rm full}$ :
 $\phi_s = - {\rm arg} M_{12}^{\rm full}$,
and thus % $\phi_{B_s}$ is the deviation from the SM prediction:
$\phi_s = 2(\beta_s - \phi_{B_s})$.
It is important to note  that the large SUSY contribution is still allowed
even though  the mass difference of  $B_s$-$\bar B_s$ \cite{Abulencia:2006ze}
is fairly consistent with the SM prediction.
This is
because the mass difference, $\Delta M_{B_s}$, can be just  twice the absolute value
of  $M_{12}^{\rm full}$.
The consistency of the mass difference between the SM prediction
and the experimental measurement just means $C_{B_s} \sim 1$,
and a large $\phi_{B_s}$ is still allowed.
For example, when $C_{B_s} \simeq 1$,
the phase $\phi_{B_s}$ is related as
$2\sin \phi_{B_s} \simeq A_{s}^{\rm NP}/A_s^{\rm SM}$,
where $A_s^{\rm NP,SM} = | M_{12}^{\rm NP,SM} |$.
In the model-independent
global analysis by the UTfit group,
the fit result is
\begin{equation}
A_s^{\rm NP}/A_s^{\rm SM} \in [0.24,1.38] \cup [1.50,2.47]
\label{UTfit}
\end{equation}
%$A_s^{\rm NP}/A_s^{\rm SM} \in [0.24,1.38] \cup [1.50,2.47]$
at 95\% probability \cite{Bona:2008jn}.
The argument of $M_{12}^{\rm NP}$ being free
in GUT models is due to the phase in off-diagonal elements in
SUSY breaking mass matrix (in the basis where $Y_d$ is a real diagonal matrix),
and one can choose an appropriate value for the new phase in the NP contribution.
Therefore, the experimental data constrains $A_s^{\rm NP}/A_s^{\rm SM}$,
and therefore, $\kappa \sin2\theta_{23}$ is constrained for a given SUSY particle spectrum.

\subsection{Box contribution}

In the MSSM with flavor universality, the chargino box diagram dominates
the SUSY contribution for $M_{12}(B_s)$.
In a  general parameter space for the soft SUSY breaking terms,
the gluino box diagram can dominate the SUSY contribution.
%the SUSY contribution of $M_{12}(B_s)$.
%
The gluino contribution can be written naively
in the mass insertion form \cite{Dutta:2006gq}
\begin{equation}
\frac{M_{12}^{\tilde g}}{M_{12}^{\rm SM}}
%\Delta M_s^{\tilde g}/\Delta M_s^{\rm SM}
\simeq
a\, [(\delta_{LL}^d)_{32}^2+ (\delta_{RR}^d)_{32}^2]
- b \, (\delta_{LL}^d)_{32} (\delta_{RR}^d)_{32},
\label{gluino-contribution}
\end{equation}
where  $a$ and $b$ depend on
squark and gluino masses, and
$\delta_{LL,RR}^d = (M^2_{\tilde d})_{LL,RR}/\tilde m^2$
($\tilde m$ is an averaged squark mass).
The mass matrix $M^2_{\tilde d}$ is a
down-type squark mass matrix $ (\tilde Q, \tilde U^{c\dagger})
M^2_{\tilde d} (\tilde Q^\dagger, \tilde U^c)^{\rm T} $ in the basis
where down-type quark mass matrix is real (positive) diagonal.
When squark and gluino masses are less than 1 TeV,
$a \sim O(1)$ and $b \sim O(100)$.
We also have contributions from $\delta_{LR}^d$,
but we neglect it since it is suppressed by $(m_b/m_{\rm SUSY})^2$.

Due to the fact that $b\gg a$, the gluino contribution
is enhanced if both left- and right-handed squark mass matrices
have off-diagonal elements.
%
%When the 23 elements of the both left- and right-handed down-type squark mass
%matrix are generated, the box diagram of $B_s$-$\bar B_s$ mixing is enhanced \cite{Dutta:2006gq}.
Therefore,
it is expected that the SUSY contribution to the $B_s$-$\bar B_s$ mixing amplitude
is large for the SO(10) model with type II seesaw~\cite{Dutta:2008xg}.

\subsection{Higgs penguin contribution and $B_s \to \mu\mu$}

The box diagram does not depend on $\tan \beta$ (ratio of the vacuum expectation values
of two Higgs fields) explicitly.
However, the flavor changing Higgs interaction (through so-called Higgs penguin diagram)
directly depend on the $\tan\beta$,
and the Higgs penguin contribution can become more important
than the box diagram when $\tan\beta$ is large \cite{Hamzaoui:1998nu,Buras:2001mb}.

The Higgs penguin contribution originates from the finite correction
of the down-type quark mass.
The effective Yukawa coupling is given as
\begin{equation}
{\cal L}^{\rm eff} = Y_d Q D^c H_d + \epsilon Q D^c H_u^*.
\end{equation}
The second term is a non-holomorphic term, which can arise from the finite correction
due to the SUSY breaking.
The effective down-type quark mass matrix is $M_d = Y_d v_d + \epsilon v_u$.
In the basis where the effective mass matrix is flavor diagonal,
flavor changing Higgs interaction can be written as
\begin{equation}
\epsilon Q D^c H_u^* - \epsilon \frac{v_u}{v_d} Q D^c H_d.
\end{equation}
Therefore, the flavor changing Higgs penguin coupling is proportional to the
finite mass correction of the down-type quark mass matrix.
The finite coupling $\epsilon$ is naively proportional to $\tan\beta$,
and thus, the dominant flavor changing Higgs interaction (second term)
is proportional to $\tan^2\beta$.
Since the $B_s$-$\bar B_s$ mixing can be generated from a double penguin diagram,
 the mixing amplitude is proportional to $\tan^4\beta$.

The effective flavor changing Higgs couplings are written as
\begin{equation}
X_{RL}^{Sij} (\bar d_i P_R d_j) S^0 + X_{LR}^{Sij} (\bar d_i P_L d_j) S^0,
\end{equation}
where $S^0$ represents for the neutral Higgs fields, $S = [H,h,A]$,
where $H$ and $h$ stand for heavier and lighter CP even neutral Higgs fields,
and $A$ is a CP odd neutral Higgs field (pseudo Higgs field).
The couplings are
\begin{eqnarray}
X_{RL}^{Sij} &=& \epsilon_{ij} \frac{1}{\sqrt2 \cos\beta} [\sin(\alpha-\beta),\cos(\alpha-\beta),-i], \\
X_{LR}^{Sij} &=& \epsilon_{ji} \frac{1}{\sqrt2 \cos\beta} [\sin(\alpha-\beta),\cos(\alpha-\beta),i],
\end{eqnarray}
where $\alpha$ is a mixing angle for $h$ and $H$.
The $B_s$-$\bar B_s$ mixing can be generated from the double left-handed penguin
(which can be generated even in the universal SUSY breaking).
However, it is proportional to the factor
\begin{equation}
\frac{\sin^2(\alpha-\beta)}{m_{H}^2} + \frac{\cos^2(\alpha-\beta)}{m_{h}^2} - \frac{1}{m_{A}^2},
\end{equation}
and the factor is almost zero since $\cos(\alpha-\beta)\simeq 0$ and
$m_{A} \simeq m_{H}$
 when $m_{A}> M_Z$ and $\tan\beta\gg1$.
In the same reason, the double right-handed penguin contribution is negligible.
On the other hand, the double penguin diagram including both left- and right-handed
Higgs penguin which is proportional to the factor
\begin{equation}
\frac{\sin^2(\alpha-\beta)}{m_{H}^2} + \frac{\cos^2(\alpha-\beta)}{m_{h}^2} + \frac{1}{m_{A}^2},
\end{equation}
and the double penguin contribution is naively proportional to
$X_{RL}^{23} X_{LR}^{23}/m_{A}^2$.
In the flavor universal SUSY breaking, the right-handed penguin coupling $X_{RL}^{23}$ is tiny,
and the double penguin contribution can not be sizable
even for a  large $\tan\beta$.
However, when the right-handed mixing is generated in the SUSY GUT models,
the double penguin diagram can be sizable for large $\tan\beta$.
We note that if there is a FCNC source in the right-handed squark mass matrix,
we do not need the off-diagonal elements in the left-handed squark mass matrix
in order to generate the sizable double penguin contribution.
Therefore, even in the minimal-type of SU(5) model,
the double penguin contribution can be sizable when $\tan\beta$ is large.
When the off-diagonal elements of left-handed squark mass matrix are generated,
the left-handed flavor changing contribution to different processes can be modified.

In the case where $\tan\beta$ is about 10,
the box contribution is dominant, and the contribution can be enhanced
in the case of SO(10) model with type II seesaw, and
a sizable contribution is not expected in the case of the minimal-type SU(5) GUT
with type I seesaw.
However, when $\tan\beta$ is around 30 or more,
the double penguin contribution can be sizable,
even in the SU(5) GUT model.
The difference between the SU(5) and the SO(10) models will not be so significant
if  only a large $B_s$-$\bar B_s$ mixing phase is observed.
In order to distinguish these models, we, however,  need to probe  other flavor changing effects,
such as $B_s \to \mu \mu$, $b \to s\gamma$ and $\tau \to \mu\gamma$.

The $B_s \to \mu\mu$ decay can be generated by a single Higgs penguin diagram \cite{Choudhury:1998ze,Buras:2001mb}.
The decay amplitude is proportional to the muon Yukawa coupling, and
thus the amplitude is proportional to $\tan^3 \beta$.
Therefore, the branching ratio is proportional to $\tan^6\beta$.
Since it can be generated by a single penguin, this decay occurs
even in the universal SUSY breaking model like the mSUGRA (minimal supergravity) \cite{Dedes:2001fv}.
The current bound of the branching ratio is Br($B_s \to \mu\mu$) = $4.7 \times 10^{-8}$~\cite{Aaltonen:2007kv}.
When $\tan\beta$ is large, this bound gives an important constraint to the parameter space
\cite{Parry:2005fp,Foster:2004vp}.
In other words, one would  expect that the $B_s \to \mu\mu$ decay will be observed very soon.

\subsection{$b\to s\gamma$ constraint}

Another important constraint for the $bs$ flavor violation is given by $b\to s\gamma$ decay \cite{Barberio:2008fa}:
\begin{equation}
{\rm Br}(b\to s\gamma) = (3.55\pm 0.26) \times 10^{-4}.
\end{equation}
The $b\to s\gamma$ decay can be generated even in the standard model,
and the NNLO determination of the branching ratio
is $(3.15\pm0.23)\times 10^{-4}$ \cite{Barberio:2008fa},
which constrains the parameter space of the MSSM.
We will choose a parameter region
to make the branching ratio to be between 2.2 to 4.2 ($\times 10^{-4}$)
in 1-loop.
In the MSSM, the chargino, gluino, and charged-Higgs contribution will be
important to the amplitude.
In the mSUGRA model, the chargino contribution will be dominant,
and low gaugino masses are excluded especially for large $\tan\beta$,
because the amplitude is proportional to $\tan\beta$.
When the charged Higgs field is light, it will generate a positive contribution
to the amplitude.
Since the chargino contribution will be negative when the Higgsino mass $\mu$
is positive, the positive $\mu$ has a wider parameter region in the MSSM.
When there is no FCNC source in SUSY breaking, the gluino contribution is tiny.
However, if there is a FCNC source in the right-handed down-type squark mass matrix,
the right-handed operator (often called $C_{7R},C_{8R}$) due to the gluino loop
can become large when $\mu$ is large.
When the left-handed squark mass matrix has off-diagonal elements,
the chargino contribution for the left-handed operator (often called $C_{7L}, C_{8L}$)
can be modified from the mSUGRA case.
It is hard to describe the allowed region in a general parameter space
with flavor violation,
but here, we give a simple correlation between the left-handed operator
for $b\to s\gamma$ and the left-handed Higgs penguin contribution.
As we have described,
the Higgs penguin contribution comes from the finite mass correction.
The $b\to s \gamma$ diagram is one-loop diagram with a photon emission.
As a result, the signs of the
contribution from the left-handed flavor violation in the squark masses
are related for the $b\to s\gamma$ contribution and
the left-handed Higgs penguin contribution.
This gives a correlation between  $b\to s\gamma$ and $B_s \to \mu\mu$.
If $\mu$ is positive, the branching ratio of $B_s\to \mu\mu$
is enhanced when the SUSY contribution arises from the flavor violating terms
cancels the chargino contribution in the mSUGRA.

\subsection{$\tau\to\mu\gamma$ constraint}

The current experimental bound of the branching ratio of $\tau\to\mu\gamma$ decay
is \cite{Hayasaka:2007vc}
\begin{equation}
{\rm Br}(\tau\to\mu\gamma) = 4.5 \times 10^{-8}.
\end{equation}

When the lepton flavor violation is correlated to the flavor violation
in the right-handed down-type squark as in the minimal-type of SU(5) model,
the $\tau\to \mu\gamma$ decay will give us the most important constraint
to obtain the large $B_s$-$\bar B_s$ phase \cite{Dutta:2008xg,Hisano:2008df}.
The minimal-type of SU(5) GUT model is predictive due to the correlation between the 
amount of quark and lepton flavor violation as we have noted previously.
Furthermore, the squark masses are raised  much more compared to
the slepton masses due to the gaugino loop contribution
since the gluino is heavier than the Bino and the Wino at low energy,
and thus the lepton flavor violation will be more sizable compared to the quark flavor violation.

In order to allow for a large phase in the $B_s$-$\bar B_s$ mixing in the minimal-type of SU(5) model,
a large flavor universal scalar mass (often called $m_0$) at the cutoff scale
is preferable.
The reasons are as follows.
The gaugino loop effects are flavor invisible
and they enhance the diagonal elements of the scalar mass matrices
while keeping the off-diagonal elements unchanged.
If the flavor universal
scalar masses at the cutoff scale become larger,
both Br($\tau\to\mu\gamma)$ and $\phi_{B_s}$ are suppressed.
However, Br($\tau\to\mu\gamma)$ is much more suppressed compared to $\phi_{B_s}$
for a given $\kappa \sin2\theta_{23}$
because the low energy slepton masses are sensitive to $m_0$
while the squark masses are not so sensitive
due to the gluino loop contribution to their masses.
In this case, however, it is hard to satisfy the muon $g-2$ \cite{g-2}
and the stau-neutralino co-annihilation region for the dark matter \cite{bd}.

When $\tan\beta$ is large, the $\tau\to\mu\gamma$ constraint
is relaxed for a large $B_s$-$\bar B_s$ phase,
because the double-penguin contribution to the $B_s$-$\bar B_s$ mixing is
proportional to $\tan^4\beta$ while the $\tau\to\mu\gamma$ is proportional
to $\tan^2\beta$.
However, the $B_s\to \mu\mu$ constraint becomes very severe in this case
since it is proportional to $\tan^6\beta$.

As we have noted, in the SO(10) model, on the other hand,
the suppression of lepton flavor violation is related to the selection of the symmetry breaking vacua,
and in fact, it is preferable that the quark flavor violation is sizable but the lepton flavor violation
is suppressed
\cite{Dutta:2007ai}.

\section{Numerical results}

We plot the figures when the NP/SM ratio of
the $B_s$-$\bar B_s$ amplitude is 0.5, $A_s^{\rm NP}/A_s^{\rm SM} = 0.5$,
and the absolute value of the full amplitude is same as SM amplitude, $C_{B_s} = 1$.
Under these choices, one can obtain that $|2\phi_{B_s}|$ is about 0.5 (rad).
We choose the unified gaugino mass $m_{1/2} = 500$ GeV,
and the sfermion mass $m_0 = 500$ GeV,
and the universal trilinear scalar coupling, $A_0 = 0$,
and $\tan\beta = 40$.
We consider that the SUSY breaking Higgs squared masses  $m_{H_u}^2$ and $m_{H_d}^2$ are not related to other scalar masses
in order to make $m_A$ and $\mu$ free parameters,
since these two parameters are important for Higgs penguin contribution
and the SUSY contribution of $b\to s\gamma$.
%h
The absolute values of 23 off-diagonal elements of squark mass matrix is fixed
to make the ratio $A_s^{\rm NP}/A_s^{\rm SM} = 0.5$.

\begin{figure}[tb]
 \center
 \includegraphics[viewport = 10 20 280 225,width=8cm]{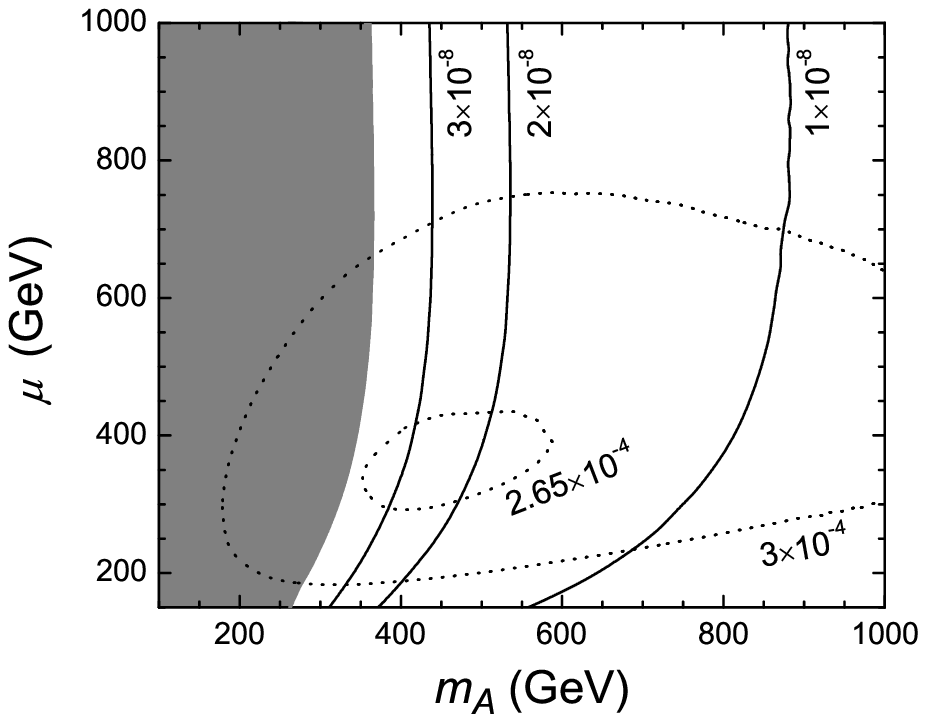}
 \includegraphics[viewport = 10 20 280 225,width=8cm]{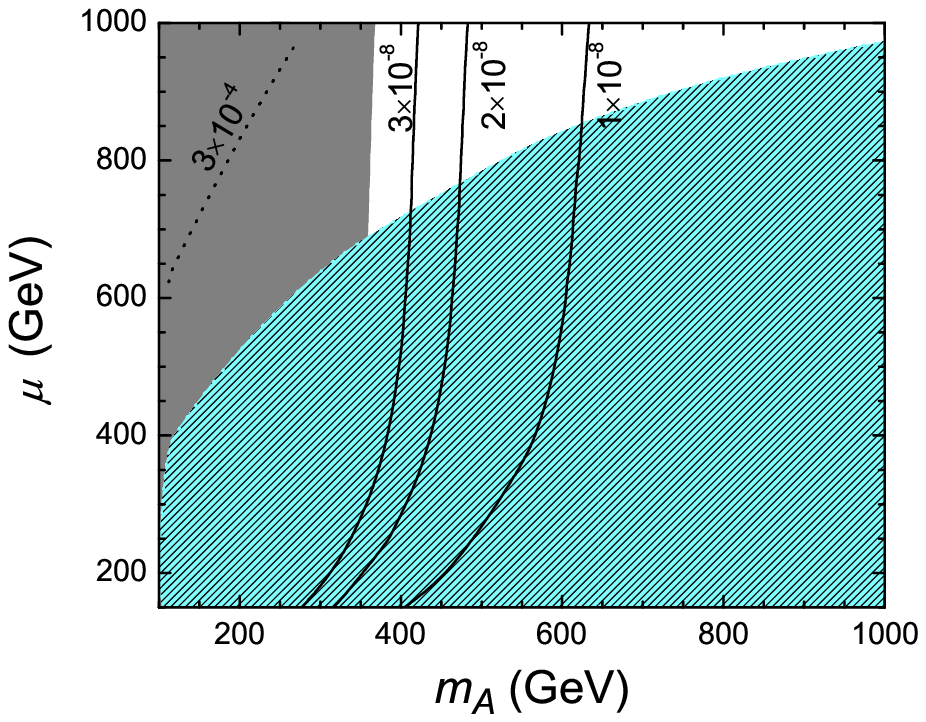}
 \caption{
Plots for the SO(10) boundary condition
when $A_s^{\rm NP}/A_s^{\rm SM} = 0.5$ and $C_{B_s} = 1$.
Solid lines show the contours of Br($B_s\to\mu\mu$).
Dot lines show the contours of Br($b\to s\gamma$).
Gray region is excluded by experimental bound of Br($B_s\to\mu\mu$).
Blue shaded region is excluded by Br($b\to s\gamma$).
Two figures are given for two signs of the 23 off-diagonal element of $M_{\tilde Q}^2$.
Details are given in the text. }
\end{figure}

In figure 1, we plot the figure in the case of the SO(10) boundary condition
with type II seesaw.
In this case, even if we fix the phase of $B_s$-$\bar B_s$ amplitude,
we still have one more phase degree of freedom.
We show the cases where the 23 off-diagonal element of left-handed squark mass matrix
is real in the basis where the down-type mass matrix is real diagonal.
(Under this choice, the modification of the left-handed penguin contribution
will be maximized.)
The other phase in the off-diagonal element in the right-handed squark mass matrix
is fixed when we choose $C_{B_s} = 1$.
The two plots in figure 1 corresponds to the two signs of the off-diagonal element
in the left-handed squark mass matrix.
Since the off-diagonal elements can have continuous phase parameter,
the two figures will morph into each other continuously by the phase degree of freedom.
In the left  plot, the Br($B_s \to \mu\mu$) is enhanced, while
it is suppressed in the right  plot.
As one can see that the Br($b\to s\gamma$) excludes most regions of the right plot,
while all this region is allowed in the left side plot,
which is explained in the previous section.
If $\mu$ is large, the gluino contributions for $C_{7R},C_{8R}$ become large
and the Br($b\to s\gamma$) constraint
is relaxed in the right  plot.
We can say that, in this parameter region,
a large $B_s$-$\bar B_s$ mixing phase can be generated by the double penguin diagram,
and the extra phase will be constrained by the Br($b\to s\gamma$) constraint.

\begin{figure}[tb]
 \center
 \includegraphics[viewport = 10 20 280 225,width=8cm]{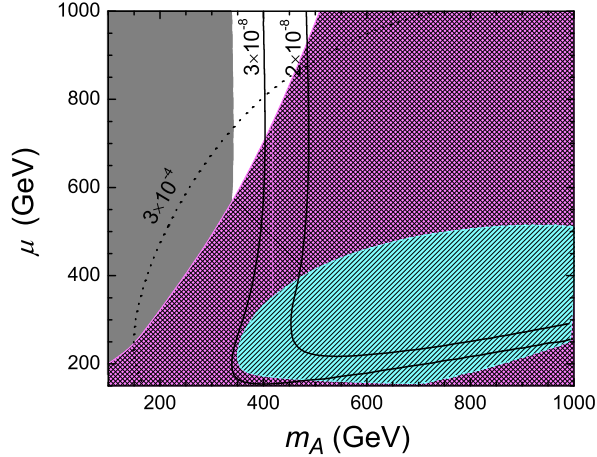}
 \caption{Plots for the SU(5) boundary condition
when $A_s^{\rm NP}/A_s^{\rm SM} = 0.5$ and $C_{B_s} = 1$.
Solid lines show the contours of Br($B_s\to\mu\mu$).
Dot lines show the contours of Br($b\to s\gamma$).
Gray region is excluded by experimental bound of Br($B_s\to\mu\mu$).
Blue shaded region is excluded by Br($b\to s\gamma$).
Pink shaded region is excluded by Br($\tau\to\mu\gamma$).
Details are given in the text.  }
\end{figure}

In figure 2, we plot the case of SU(5) boundary condition.
We choose  the $\kappa$ values to be  exactly same for $\tilde{L}$ and $\tilde{D^c}$
for simplicity.
The parameters are same as before, $m_0 = m_{1/2} = 500$ GeV, $A_0 =0$ and $\tan\beta = 40$.
In this case, the phase is fixed (up to sign) when we choose
$A_s^{\rm NP}/A_s^{\rm SM} = 0.5$ and $C_{B_s} = 1$.
One can see that, the Br($\tau\to\mu\gamma$) constraint excludes
most  regions of the plot.
In other words, Br($B_s\to \mu \mu$) has to be large enough to be detected
under this boundary condition for large $\tan\beta$.
This is because of the following reasons.
When $m_A$ is large, the double penguin contribution is suppressed.
Then, $\kappa$ has to be large in order to obtain a large $B_s$-$\bar B_s$ mixing phase.
However, a large $\kappa$ is excluded by the Br($\tau\to\mu\gamma$) constraint.
As a result, a heavy pseudo Higgs field is excluded, and thus the Br($B_s \to\mu\mu$)
has to be large.
As we have noted, the Br($\tau\to\mu\gamma$) constraint is relaxed when $m_0$ is large.

We note that using
the universal scalar mass boundary condition for sfermion and Higgs fields
($m_0 = m_{H_u} = m_{H_d}$), it is very hard to obtain the large $B_s$-$\bar B_s$ mixing phase
due to Br($\tau\to\mu\gamma$) constraint under the SU(5) boundary condition with type I seesaw,
since $m_A$ and $\mu$ is not free in this case.
Actually, $m_A$ is not so low to enhance the double penguin diagram in this universal boundary condition
as a consequence of the fact that
 the gaugino mass should be large enough to satisfy the $b\to s\gamma$ constraint especially
for large $\tan\beta$.
However, when $\tan\beta$ is about 50, the pseudo Higgs mass $m_A$
becomes lower due to bottom Yukawa contribution in RGEs.
Furthermore, as we have noted, the
double-penguin contribution to the $B_s$-$\bar B_s$ mixing is
proportional to $\tan^4\beta$ while the $\tau\to\mu\gamma$ is proportional
to $\tan^2\beta$.
Therefore, in this case, a large $B_s$-$\bar B_s$ mixing phase can survive
satisfying the $\tau\to\mu\gamma$ constraint
even in the universal scalar mass condition.
In this case, the branching ratio of $B_s \to \mu\mu$ has to be at the edge of the current bound.

\section{Conclusion}

We investigated the GUT models
when the $B_s$-$\bar B_s$ mixing phase can become really large
as indicated in the Fermilab experiments.
We considered two cases:
one is the minimal-type of SU(5) model with type I seesaw.
The other is the minimal-type of SO(10) model with type II seesaw.
The difference between the two boundary condition is whether there exists
a sizable off-diagonal element in the
left-handed squark mass matrix.
It is important to note that the sources of FCNC will be  restricted
in the GUT models
if the large phase of $B_s$-$\bar B_s$ mixing persists in the upcoming result
in the Fermilab.

For small $\tan\beta$, the SUSY contribution of $B_s$-$\bar B_s$ mixing
amplitude is dominated by the gluino box contribution,
and the phase of $B_s$-$\bar B_s$ mixing will be more enhanced
under the SO(10) boundary condition compared
the SU(5) boundary condition \cite{Dutta:2006gq,Dutta:2008xg}.
When $\tan\beta$ is large,
the double penguin contribution will dominate
in both SU(5) and SO(10) boundary condition.
Under the SO(10) boundary condition, the left-handed FCNC source will modify
the left-handed Higgs penguin as well as the $C_{7L},C_{8L}$ operators
for the Br($b\to s\gamma$) decay,
depending the phase of the 23 off-diagonal element.
When the phase of $B_s$-$\bar B_s$ mixing is large,
the phase of the 23 off-diagonal element in the left-handed squark mass matrix
is restricted especially when Higgsino mass $\mu$ is small
as shown in Fig. 1.
When the phase is suitable to satisfy the Br($b\to s\gamma$) bound,
the left-handed penguin contribution is slightly enhanced and
the Br($b\to s\gamma$) is larger compared to the case with no left-handed FCNC source.
Under the SU(5) boundary condition, the pseudo Higgs mass should be low enough
to satisfy the Br($\tau\to\mu\gamma$) constraint for a given parameter
as shown in Fig. 2,
and then the Br($B_s \to\mu\mu$) has to be sizable, and it can be detected very soon.

In this paper, we have concentrated on the importance of
the 2nd and 3rd generation FCNC effects
such as Br($\tau\to \mu\gamma$) and $\phi_{B_s}$ correlation in GUT models,
since they can be correlated directly by 23 mixing.
The constraints from Br($\mu\to e\gamma$) decay, $K$-$\bar K$ and $B_d$-$\bar B_d$ mixings,
 may be also important,
but these effects %highly
depend on the details of
 flavor structure which can have a freedom of cancellation. %\cite{Dutta:2006zt}.
%
%Therefore, we have not talked about the $\mu\to e\gamma$ constraint.
We refer to the Ref.\cite{Dutta:2006zt} for an analysis of
flavor violation including the first generation.

\section*{Acknowledgments}

This work %of B. D. and Y. M.
is supported in part by the DOE grant
DE-FG02-95ER40917.


\begin{thebibliography}{99}
%\newcommand{\wwwspires}{http://www.yukawa.kyoto-u.ac.jp/spires/find/hep/www}
%%%%%%%%%%%%%%%%%%%%%%%%%%%%%%%%%
%
%
%
%
%
%
%
%



%\cite{Aaltonen:2007he}
\bibitem{Aaltonen:2007he}
  T.~Aaltonen {\it et al.}  [CDF Collaboration],
  %``First Flavor-Tagged Determination of Bounds on Mixing-Induced CP Violation
  %in Bs -> J/psi phi Decays,''
  Phys.\ Rev.\ Lett.\  {\bf 100}, 161802 (2008)
  [arXiv:0712.2397 [hep-ex]];
  %%CITATION = PRLTA,100,161802;%%
%
%\cite{:2008fj}
%\bibitem{:2008fj}
  V.~M.~Abazov {\it et al.}  [D0 Collaboration],
  %``Measurement of $\boldmath {B_s^0}$ mixing parameters from the flavor-tagged
  %decay $B^0_s \to J/\psi \phi$,''
  Phys.\ Rev.\ Lett.\  {\bf 101}, 241801 (2008)
  [arXiv:0802.2255 [hep-ex]];
  %%CITATION = PRLTA,101,241801;%%
%
%\cite{Tonelli:2008ey}
%\bibitem{Tonelli:2008ey}
  D.~Tonelli [CDF Collaboration],
  %``Search for New Physics in the B0s mixing phase,''
  arXiv:0810.3229 [hep-ex].
  %%CITATION = ARXIV:0810.3229;%%




%\cite{Bona:2008jn}
\bibitem{Bona:2008jn}
  M.~Bona {\it et al.}  [UTfit Collaboration],
  %``First Evidence of New Physics in b <--> s Transitions,''
  arXiv:0803.0659 [hep-ph].
  %%CITATION = ARXIV:0803.0659;%%





%\cite{Gabbiani:1988rb}
\bibitem{Gabbiani:1988rb}
  F.~Gabbiani and A.~Masiero,
  %``Fcnc In Generalized Supersymmetric Theories,''
  Nucl.\ Phys.\ B {\bf 322}, 235 (1989);
  %%CITATION = NUPHA,B322,235;%%
%
%\cite{Hagelin:1992tc}
%\bibitem{Hagelin:1992tc}
  J.~Hagelin, S.~Kelley and T.~Tanaka,
  %``Supersymmetric flavor changing neutral currents: Exact amplitudes and
  %phenomenological analysis,''
  Nucl.\ Phys.\ B {\bf 415}, 293 (1994);
  %%CITATION = NUPHA,B415,293;%%
%
%\cite{Gabbiani:1996hi}
%\bibitem{Gabbiani:1996hi}
  F.~Gabbiani, % {\it et al.},
  E.~Gabrielli, A.~Masiero and L.~Silvestrini,
  %``A complete analysis of FCNC and CP constraints in general SUSY extensions
  %of the standard model,''
  Nucl.\ Phys.\ B {\bf 477}, 321 (1996)
  [hep-ph/9604387].
  %%CITATION = HEP-PH 9604387;%%






%\cite{Borzumati:1986qx}
\bibitem{Borzumati:1986qx}
  F.~Borzumati and A.~Masiero,
  %``Large Muon And Electron Number Violations In Supergravity Theories,''
  Phys.\ Rev.\ Lett.\  {\bf 57}, 961 (1986);
  %%CITATION = PRLTA,57,961;%%
%
%\cite{Hisano:1995nq}
%\bibitem{Hisano:1995nq}
  J.~Hisano, % {\it et al.},
  T.~Moroi, K.~Tobe, M.~Yamaguchi and T.~Yanagida,
  %``Lepton flavor violation in the supersymmetric standard model with seesaw
  %induced neutrino masses,''
  Phys.\ Lett.\ B {\bf 357}, 579 (1995)
  [hep-ph/9501407].
  %%CITATION = HEP-PH 9501407;%%





\bibitem{Barbieri:1994pv}
%\cite{Hall:1985dx}
%\bibitem{Hall:1985dx}
  L.~J.~Hall, V.~A.~Kostelecky and S.~Raby,
  %``New Flavor Violations In Supergravity Models,''
  Nucl.\ Phys.\ B {\bf 267}, 415 (1986);
  %%CITATION = NUPHA,B267,415;%%
%
%\cite{Barbieri:1994pv}
%\bibitem{Barbieri:1994pv}
  R.~Barbieri and L.~J.~Hall,
  %``Signals for supersymmetric unification,''
  Phys.\ Lett.\ B {\bf 338}, 212 (1994)
  [hep-ph/9408406];
  %%CITATION = HEP-PH 9408406;%%
%
%\cite{Hisano:1996qq}
%\bibitem{Hisano:1996qq}
  J.~Hisano, T.~Moroi, K.~Tobe and M.~Yamaguchi,
  %``Exact event rates of lepton flavor violating processes in  supersymmetric
  %SU(5) model,''
  Phys.\ Lett.\ B {\bf 391}, 341 (1997)
%  [Erratum-ibid.\ B {\bf 397}, 357 (1997)]
  [hep-ph/9605296].
  %%CITATION = HEP-PH 9605296;%%




%\cite{Moroi:2000tk}
\bibitem{Moroi:2000tk}
  T.~Moroi,
  %``CP violation in B/d --> Phi K(S) in SUSY GUT with right-handed
  %neutrinos,''
  Phys.\ Lett.\  B {\bf 493}, 366 (2000)
  [hep-ph/0007328].
  %%CITATION = PHLTA,B493,366;%%



%\cite{Dutta:2006gq}
\bibitem{Dutta:2006gq}
  B.~Dutta and Y.~Mimura,
  %``B/s - anti-B/s mixing in grand unified models,''
  Phys.\ Rev.\ Lett.\  {\bf 97}, 241802 (2006)
  [hep-ph/0607147].
  %%CITATION = PRLTA,97,241802;%%



%\cite{Parry:2005fp}
\bibitem{Parry:2005fp}
  J.~K.~Parry,
  %``Lepton flavor violating Higgs boson decays, tau --> mu gamma, B/s -->  mu+
  %mu- and theta(13) in the constrained MSSM with massive neutrinos  and large
  %tan beta,''
  Nucl.\ Phys.\  B {\bf 760}, 38 (2007)
  [hep-ph/0510305];
  %%CITATION = NUPHA,B760,38;%%
%
%\cite{Parry:2006vq}
%\bibitem{Parry:2006vq}
%  J.~K.~Parry,
  %``B/s0 anti-B/s0 mixing in the MSSM with large tan(beta),''
  Mod.\ Phys.\ Lett.\  A {\bf 21}, 2853 (2006)
  [hep-ph/0608192];
  %%CITATION = MPLAE,A21,2853;%%
%
%\cite{Parry:2007fe}
%\bibitem{Parry:2007fe}
  J.~K.~Parry and H.~h.~Zhang,
  %``B(d,s)-anti-B(d,s) mixing and Lepton Flavour Violation in SUSY GUTs: impact
  %of the first measurements of phi(s),''
  Nucl.\ Phys.\  B {\bf 802}, 63 (2008)
  [arXiv:0710.5443 [hep-ph]].
  %%CITATION = NUPHA,B802,63;%%


%\cite{Dutta:2008xg}
\bibitem{Dutta:2008xg}
  B.~Dutta and Y.~Mimura,
  %``Large Phase of B_s-\bar B_s Mixing in Supersymmetric Grand Unified
  %Theories,''
  Phys.\ Rev.\  D {\bf 78}, 071702 (2008)
  [arXiv:0805.2988 [hep-ph]].
  %%CITATION = PHRVA,D78,071702;%%



%\cite{Hisano:2008df}
\bibitem{Hisano:2008df}
  J.~Hisano and Y.~Shimizu,
  %``CP Violation in B_s Mixing in the SUSY SU(5) GUT with Right-handed
  %Neutrinos,''
  Phys.\ Lett.\  B {\bf 669}, 301 (2008)
  [arXiv:0805.3327 [hep-ph]];
  %%CITATION = PHLTA,B669,301;%%
%
%\cite{Park:2008vv}
%\bibitem{Park:2008vv}
  J.~h.~Park and M.~Yamaguchi,
  %``B_s mixing phase and lepton flavor violation in supersymmetric SU(5),''
  Phys.\ Lett.\  B {\bf 670}, 356 (2009)
  [arXiv:0809.2614 [hep-ph]].
  %%CITATION = PHLTA,B670,356;%%








%\cite{Hamzaoui:1998nu}
\bibitem{Hamzaoui:1998nu}
  C.~Hamzaoui, M.~Pospelov and M.~Toharia,
  %``Higgs-mediated FCNC in supersymmetric models with large tan(beta),''
  Phys.\ Rev.\  D {\bf 59}, 095005 (1999)
  [hep-ph/9807350];
  %%CITATION = PHRVA,D59,095005;%%
%
%\cite{Gorbahn:2009pp}
%\bibitem{Gorbahn:2009pp}
  M.~Gorbahn, S.~Jager, U.~Nierste and S.~Trine,
  %``The supersymmetric Higgs sector and B-Bbar mixing for large tan beta,''
  arXiv:0901.2065 [hep-ph].
  %%CITATION = ARXIV:0901.2065;%%



%\cite{Buras:2001mb}
\bibitem{Buras:2001mb}
  A.~J.~Buras, P.~H.~Chankowski, J.~Rosiek and L.~Slawianowska,
  %``Delta(M(s))/Delta(M(d)), sin 2beta and the angle gamma in the presence  of
  %new Delta(F) = 2 operators,''
  Nucl.\ Phys.\  B {\bf 619}, 434 (2001)
  [hep-ph/0107048];
  %%CITATION = NUPHA,B619,434;%%
%
%\cite{Buras:2002wq}
%\bibitem{Buras:2002wq}
%  A.~J.~Buras, P.~H.~Chankowski, J.~Rosiek and L.~Slawianowska,
  %``Correlation between Delta M(s) and B/(s,d)0 --> mu+ mu- in  supersymmetry
  %at large tan(beta),''
  Phys.\ Lett.\  B {\bf 546}, 96 (2002)
  [hep-ph/0207241];
  %%CITATION = PHLTA,B546,96;%%
%
%\cite{Buras:2002vd}
%\bibitem{Buras:2002vd}
%  A.~J.~Buras, P.~H.~Chankowski, J.~Rosiek and L.~Slawianowska,
  %``Delta(M(d,s)), B/(d,s)0 --> mu+ mu- and B --> X/s gamma in supersymmetry at
  %large tan(beta),''
  Nucl.\ Phys.\  B {\bf 659}, 3 (2003)
  [hep-ph/0210145].
  %%CITATION = NUPHA,B659,3;%%





%\cite{Choudhury:1998ze}
\bibitem{Choudhury:1998ze}
  S.~R.~Choudhury and N.~Gaur,
  %``Dileptonic decay of B/s meson in SUSY models with large tan(beta),''
  Phys.\ Lett.\  B {\bf 451}, 86 (1999)
  [hep-ph/9810307];
  %%CITATION = PHLTA,B451,86;%%
%
%\cite{Babu:1999hn}
%\bibitem{Babu:1999hn}
  K.~S.~Babu and C.~F.~Kolda,
  %``Higgs-mediated B0 --> mu+ mu- in minimal supersymmetry,''
  Phys.\ Rev.\ Lett.\  {\bf 84}, 228 (2000)
  [hep-ph/9909476];
  %%CITATION = PRLTA,84,228;%%
%
%\cite{Huang:2000sm}
%\bibitem{Huang:2000sm}
  C.~S.~Huang, W.~Liao, Q.~S.~Yan and S.~H.~Zhu,
  %``B/s --> l+ l- in a general 2HDM and MSSM,''
  Phys.\ Rev.\  D {\bf 63}, 114021 (2001)
  [Erratum-ibid.\  D {\bf 64}, 059902 (2001)]
  [hep-ph/0006250].
  %%CITATION = PHRVA,D63,114021;%%
%
%\cite{Chankowski:2000ng}
%\bibitem{Chankowski:2000ng}
  P.~H.~Chankowski and L.~Slawianowska,
  %``B0/d,s --> mu- mu+ decay in the MSSM,''
  Phys.\ Rev.\  D {\bf 63}, 054012 (2001)
  [hep-ph/0008046];
  %%CITATION = PHRVA,D63,054012;%%
%
%\cite{Bobeth:2001sq}
%\bibitem{Bobeth:2001sq}
  C.~Bobeth, T.~Ewerth, F.~Kruger and J.~Urban,
  %``Analysis of neutral Higgs-boson contributions to the decays anti-B/s  -->
  %l+ l- and anti-B --> K l+ l-,''
  Phys.\ Rev.\  D {\bf 64}, 074014 (2001)
  [hep-ph/0104284];
  %%CITATION = PHRVA,D64,074014;%%
%
%\cite{Bobeth:2002ch}
%\bibitem{Bobeth:2002ch}
%  C.~Bobeth, T.~Ewerth, F.~Kruger and J.~Urban,
  %``Enhancement of B(anti-B/d --> mu+ mu-)/B(anti-B/s --> mu+ mu-)  in the MSSM
  %with minimal flavour violation and large tan beta,''
  Phys.\ Rev.\  D {\bf 66}, 074021 (2002)
  [hep-ph/0204225].
  %%CITATION = PHRVA,D66,074021;%%



%\cite{Foster:2004vp}
\bibitem{Foster:2004vp}
  J.~Foster, K.~Okumura and L.~Roszkowski,
  %``New Higgs effects in B physics in supersymmetry with general flavour
  %mixing,''
  Phys.\ Lett.\  B {\bf 609}, 102 (2005)
  [hep-ph/0410323];
  %%CITATION = PHLTA,B609,102;%%
%
%\cite{Foster:2005wb}
%\bibitem{Foster:2005wb}
%  J.~Foster, K.~Okumura and L.~Roszkowski,
  %``Probing the flavour structure of supersymmetry breaking with rare
  %B-processes: A beyond leading order analysis,''
  JHEP {\bf 0508}, 094 (2005)
  [hep-ph/0506146];
  %%CITATION = JHEPA,0508,094;%%
%
%\cite{Foster:2006ze}
%\bibitem{Foster:2006ze}
%  J.~Foster, K.~Okumura and L.~Roszkowski,
  %``New constraints on SUSY flavour mixing in light of recent measurements at
  %the Tevatron,''
  Phys.\ Lett.\  B {\bf 641}, 452 (2006)
  [hep-ph/0604121].
  %%CITATION = PHLTA,B641,452;%%




%\cite{Minkowski:1977sc}
\bibitem{Minkowski:1977sc}
  P.~Minkowski,
  %``Mu $\to$ E Gamma At A Rate Of One Out Of 1-Billion Muon Decays?,''
  Phys.\ Lett.\ B {\bf 67}, 421 (1977);
  %%CITATION = PHLTA,B67,421;%%
%
T. Yanagida, in proc. of KEK workshop, eds. O. Sawada and S. Sugamoto
(Tsukuba, 1979);
%
M. Gell-Mann, P. Ramond and R. Slansky,
in {\it Supergravity}, eds. P. van Nieuwenhuizen and D.~Z. Freedman
(North-Holland, Amsterdam, 1979);
%
%\cite{Mohapatra:1979ia}
%\bibitem{Mohapatra:1979ia}
  R.~N.~Mohapatra and G.~Senjanovi\'c,
  %``Neutrino Mass And Spontaneous Parity Nonconservation,''
  Phys.\ Rev.\ Lett.\  {\bf 44}, 912 (1980).
  %%CITATION = PRLTA,44,912;%%





%\cite{Schechter:1980gr}
\bibitem{Schechter:1980gr}
  J.~Schechter and J.~W.~F.~Valle,
  %``Neutrino Masses In SU(2) X U(1) Theories,''
  Phys.\ Rev.\ D {\bf 22}, 2227 (1980);
  %%CITATION = PHRVA,D22,2227;%%
%
%\cite{Mohapatra:1980yp}
%\bibitem{Mohapatra:1980yp}
  R.~N.~Mohapatra and G.~Senjanovic,
  %``Neutrino Masses And Mixings In Gauge Models With Spontaneous Parity
  %Violation,''
  Phys.\ Rev.\ D {\bf 23}, 165 (1981);
  %%CITATION = PHRVA,D23,165;%%
%
%\cite{Lazarides:1980nt}
%\bibitem{Lazarides:1980nt}
  G.~Lazarides, Q.~Shafi and C.~Wetterich,
  %``Proton Lifetime And Fermion Masses In An SO(10) Model,''
  Nucl.\ Phys.\ B {\bf 181}, 287 (1981).
  %%CITATION = NUPHA,B181,287;%%






%\cite{Dutta:2006zt}
\bibitem{Dutta:2006zt}
  B.~Dutta and Y.~Mimura,
  %``Modification of the unitarity relation for sin(2beta)-V(ub) in
  %supersymmetric models,''
  Phys.\ Rev.\  D {\bf 75}, 015006 (2007)
  [hep-ph/0611268];
  %%CITATION = PHRVA,D75,015006;%%
%
%\cite{Dutta:2007ue}
%\bibitem{Dutta:2007ue}
%  B.~Dutta and Y.~Mimura,
  %``Constraint from D - anti-D Mixing in Left-Right Symmetric Models,''
  Phys.\ Rev.\  D {\bf 77}, 051701 (2008)
  [arXiv:0708.3080 [hep-ph]].
  %%CITATION = PHRVA,D77,051701;%%








%\cite{Babu:1992ia}
\bibitem{Babu:1992ia}
  K.~S.~Babu and R.~N.~Mohapatra,
  %``Predictive Neutrino Spectrum In Minimal SO(10) Grand Unification,''
  Phys.\ Rev.\ Lett.\  {\bf 70}, 2845 (1993)
  [hep-ph/9209215].
  %%CITATION = PRLTA,70,2845;%%





%\cite{Dutta:2004wv}
\bibitem{Dutta:2004wv}
  B.~Dutta, Y.~Mimura and R.~N.~Mohapatra,
  %``CKM CP violation in a minimal SO(10) model for neutrinos and its
  %implications,''
  Phys.\ Rev.\  D {\bf 69}, 115014 (2004)
  [hep-ph/0402113];
  %%CITATION = PHRVA,D69,115014;%%
%
%\cite{Dutta:2004hp}
%\bibitem{Dutta:2004hp}
%  B.~Dutta, Y.~Mimura and R.~N.~Mohapatra,
  %``Neutrino masses and mixings in a predictive SO(10) model with CKM CP
  %violation,''
  Phys.\ Lett.\ B {\bf 603}, 35 (2004)
  [hep-ph/0406262];
  %%CITATION = HEP-PH 0406262;%%
%
%\cite{Dutta:2004zh}
%\bibitem{Dutta:2004zh}
%  B.~Dutta, Y.~Mimura and R.~N.~Mohapatra,
  %``Suppressing proton decay in the minimal SO(10) model,''
  Phys.\ Rev.\ Lett.\  {\bf 94}, 091804 (2005)
  [hep-ph/0412105];
  %%CITATION = PRLTA,94,091804;%%
%
%\cite{Dutta:2005ni}
%\bibitem{Dutta:2005ni}
%  B.~Dutta, Y.~Mimura and R.~N.~Mohapatra,
  %``Neutrino mixing predictions of a minimal SO(10) model with suppressed
  %proton decay,''
  Phys.\ Rev.\  D {\bf 72}, 075009 (2005)
  [hep-ph/0507319].
  %%CITATION = PHRVA,D72,075009;%%






%\cite{Dutta:2007ai}
\bibitem{Dutta:2007ai}
  B.~Dutta, Y.~Mimura and R.~N.~Mohapatra,
  %``Proton Decay and Flavor Violating Thresholds in SO(10) Models,''
  Phys.\ Rev.\ Lett.\  {\bf 100}, 181801 (2008)
  [arXiv:0712.1206 [hep-ph]].
  %%CITATION = PRLTA,100,181801;%%



\bibitem{Abulencia:2006ze}
%\cite{Abazov:2006dm}
%\bibitem{Abazov:2006dm}
  V.~Abazov {\it et al.}  [D0],
  %``First direct two-sided bound on the B/s0 oscillation frequency,''
  Phys.\ Rev.\ Lett.\  {\bf 97}, 021802 (2006)
  [hep-ex/0603029];
  %%CITATION = HEP-EX 0603029;%%
%
%CDF Collaboration,
%\url{http://www-cdf.fnal.gov/physics/new/bottom/060406.blessed-Bsmix/}.
%\cite{unknown:2006mq}
%\bibitem{unknown:2006mq}
  A.~Abulencia {\it et al.} [CDF],
  %``Measurement of the B/s0 anti-B/s0 oscillation frequency,''
  Phys.\ Rev.\ Lett.\  {\bf 97}, 062003 (2006)
  [hep-ex/0606027].
  %%CITATION = HEP-EX 0606027;%%


%\cite{Dedes:2001fv}
\bibitem{Dedes:2001fv}
  A.~Dedes, H.~K.~Dreiner and U.~Nierste,
  %``Correlation of B/s --> mu+ mu- and (g-2)(mu) in minimal supergravity,''
  Phys.\ Rev.\ Lett.\  {\bf 87}, 251804 (2001)
  [hep-ph/0108037];
  %%CITATION = PRLTA,87,251804;%%
%
%\cite{Arnowitt:2002cq}
%\bibitem{Arnowitt:2002cq}
  R.~L.~Arnowitt, B.~Dutta, T.~Kamon and M.~Tanaka,
  %``Detection of B/s --> mu+ mu- at the Tevatron Run II and constraints on  the
  %SUSY parameter space,''
  Phys.\ Lett.\  B {\bf 538}, 121 (2002)
  [hep-ph/0203069];
  %%CITATION = PHLTA,B538,121;%%
%
J.~K.~Mizukoshi, X.~Tata and Y.~Wang,
  %``Higgs-mediated leptonic decays of B/s and B/d mesons as probes of
  %supersymmetry,''
  Phys.\ Rev.\  D {\bf 66}, 115003 (2002)
  [hep-ph/0208078];
  %%CITATION = PHRVA,D66,115003;%%
%
   J.~R.~Ellis, K.~A.~Olive and V.~C.~Spanos,
  %``On the interpretation of B/s --> mu+ mu- in the CMSSM,''
  Phys.\ Lett.\  B {\bf 624}, 47 (2005)
  [hep-ph/0504196].
  %%CITATION = PHLTA,B624,47;%%




%\cite{:2007kv}
\bibitem{Aaltonen:2007kv}
  T.~Aaltonen {\it et al.}  [CDF Collaboration],
  %``Search for $B_s \to \mu^+\mu^-$ and $B_d \to \mu^+\mu^-$ Decays with
  %2fb$^{-1}$ of $p\bar{p}$ Collisions,''
  Phys.\ Rev.\ Lett.\  {\bf 100}, 101802 (2008)
  [arXiv:0712.1708 [hep-ex]].
  %%CITATION = PRLTA,100,101802;%%






%\cite{Barberio:2008fa}
\bibitem{Barberio:2008fa}
  E.~Barberio {\it et al.}  [Heavy Flavor Averaging Group],
  %``Averages of b-hadron and c-hadron Properties at the End of 2007,''
  arXiv:0808.1297 [hep-ex].
  %%CITATION = ARXIV:0808.1297;%%




%\cite{Hayasaka:2007vc}
\bibitem{Hayasaka:2007vc}
  K.~Hayasaka {\it et al.}  [Belle Collaboration],
  %``New search for tau --> mu gamma and tau --> e gamma decays at Belle,''
  Phys.\ Lett.\  B {\bf 666}, 16 (2008)
  [arXiv:0705.0650 [hep-ex]];
  %%CITATION = PHLTA,B666,16;%%
%
%\cite{Aubert:2005ye}
%\bibitem{Aubert:2005ye}
  B.~Aubert {\it et al.}  [BABAR Collaboration],
%   ``Search for lepton flavor violation in the decay tau $\to$ mu gamma,''
  %
  Phys.\ Rev.\ Lett.\  {\bf 95}, 041802 (2005)
  [hep-ex/0502032].
  %%CITATION = HEP-EX 0502032;%%








\bibitem{g-2}
Muon $g-2$ Collaboration, G.~W.~Bennett {\it et al.},
  %``Measurement of the negative muon anomalous magnetic moment to 0.7-ppm,''
  Phys. Rev. Lett. {\bf 92}, 161802 (2004)
  [hep-ex/0401008].
%%CITATION = PRLTA,92,161802;%%;






\bibitem{bd}
V.~Khotilovich, R.~Arnowitt, B.~Dutta and T.~Kamon,
  %``The stau neutralino co-annihilation region at an international linear
  %collider,''
  Phys.\ Lett.\  B {\bf 618}, 182 (2005)
  [hep-ph/0503165].
  %%CITATION = PHLTA,B618,182;%%























\end{thebibliography}
\end{document}